# Beyond Score Accuracy: Examining the Diagnostic Quality of LLM-Generated Structured Assessment in Higher Education

**Xi Zhao[a*], Xinyue Jiao[b] & Zhen Xu[a]**
[a] *Teachers College, Columbia University, USA*
[b] *School of Culture, Education, and Human Development, New York University, USA*
*xz3124@tc.columbia.edu

**Abstract:** As Large Language Models (LLMs) are increasingly adopted for automated grading and feedback in higher education, their structured outputs, including multi-dimensional rubric scores, detailed feedback comments, and improvement suggestions, create an appearance of thorough analytic evaluation. This study examines whether these outputs deliver what they appear to offer. Using the JorGPT dataset of 3,041 student responses to 50 open-ended computer science questions, scored by both human instructors and three commercial LLMs, we identify three systematic discrepancies between the apparent and actual quality of LLM-generated grading and feedback. The sub-dimension scores are highly correlated ($r = 0.82–0.99$, VIF up to 45), providing redundant rather than independent diagnostic information. The textual feedback rarely detects student misconceptions (5–7% vs. 15–31% for teachers), functioning as a coverage checklist rather than a diagnostic instrument. The feedback tone remains uniformly positive regardless of response quality, lacking the severity modulation observed in human feedback. Additionally, grading accuracy varies significantly by knowledge domain, with procedural topics most reliable. These findings provide empirically grounded guidance on which aspects of LLM-generated grading and feedback can be relied upon and which require continued human oversight.

## 1. Introduction

Large Language Models (LLMs) have demonstrated considerable potential in educational assessment, prompting growing interest in their use as automated grading and feedback tools in higher education. Recent systematic reviews have documented a rapid increase in AI-assisted text-based assessment systems across post-secondary contexts (Gao et al., 2024), and researchers have begun to examine the role of LLMs such as ChatGPT in automating grading processes (Kooli & Yusuf, 2025). Empirical studies have reported promising score-level alignment between LLM-generated and human grades. For instance, Khalf et al. (2026) found that fine-tuned transformer models achieved a Pearson correlation of 0.88 with human scores on short-answer grading tasks. Mendonça et al. (2025) compared LLaMA 3.2 and GPT-4o against human evaluators for scoring open-ended programming questions, finding that the premium model achieved statistically equivalent grading patterns to human raters, particularly for code-based questions.

Beyond scoring accuracy, a growing body of work has investigated how LLM-generated feedback compares with human feedback. Usher (2025) compared feedback from AI chatbots, peers, and instructors in a higher education course, finding that chatbot-generated feedback sometimes included irrelevant or incorrect suggestions requiring critical evaluation. Alyahyan et al. (2025) mapped the emerging literature on LLMs as feedback providers in higher education, highlighting open questions about their effectiveness and reliability. Liu et al. (2025) further demonstrated that LLM-generated responses in online learning contexts exhibit sentiment bias across demographic attributes, raising concerns about the fairness and consistency of automated feedback.

However, effective formative feedback requires more than factual accuracy. A recent meta-analysis confirmed that the impact of feedback on student learning is substantially influenced by its information content, with feedback addressing the nature of errors yielding stronger effects than feedback merely indicating correctness (Wisniewski et al., 2020). In computer science education, programming misconceptions have been extensively documented and shown to persist when feedback fails to address underlying conceptual errors (Qian & Lehman, 2017). Beyond diagnostic content, the tone and severity of feedback also shape how students interpret and act upon it (Hattie & Timperley, 2007). These qualities, including diagnostic specificity, misconception identification, and tonal calibration, have received limited empirical attention in the context of LLM-generated structured grading and feedback.

Modern LLM-based assessment systems can produce structured outputs that go beyond a single score, including multiple rubric sub-dimension scores, composite totals, and detailed textual feedback. Such outputs bear a surface resemblance to the analytic assessments traditionally produced by expert human graders, in which independent rubric dimensions are designed to provide distinct diagnostic information (Jonsson & Svingby, 2007). This resemblance may lead educators to assume that LLM-generated structured outputs provide similarly independent and diagnostically meaningful information. However, this assumption has received limited empirical scrutiny. To address these gaps, this study examines the quality of LLM-generated grading and feedback in an undergraduate computer science course. Specifically, we investigate the following research questions:

**RQ1:** To what extent do LLM-generated grades align with human instructor grades, and does this alignment vary across knowledge domains?

**RQ2:** Do the sub-dimension scores produced by LLMs provide independent diagnostic information, or do they reflect a single underlying construct?

**RQ3:** How does the content and tone of LLM-generated feedback differ from that of human instructors?

## 2. Data and Method

### *2.1 Dataset*

This study uses the JorGPT dataset (Sánchez-Soriano et al., 2026), a publicly available multilingual collection of 3,041 authentic student answers to 50 open-ended computer science questions, collected from real university assessments during the 2025–2026 academic year at the Universidad Francisco de Vitoria, Spain. The data collection was conducted under a protocol reviewed and approved by the Research Ethics Committee of Universidad Francisco de Vitoria (Project identification code: 31/2026), in accordance with the Declaration of Helsinki. Informed consent was obtained from all participants, and all student responses were anonymized using irreversible identifiers prior to public release. The dataset originates from undergraduate courses in computer science and software engineering, covering topics such as software development methodologies, software testing strategies, and fundamental programming concepts. All student responses were produced under regular assessment conditions, ensuring ecological validity.

Each student response is accompanied by a human instructor evaluation consisting of an integer grade on a 0–10 scale and a qualitative feedback comment. The instructor graded student answers anonymously within the learning management system to prevent potential bias. In addition, each response was independently evaluated by three commercial LLMs (DeepSeek-chat-V3.2, Qwen-flash-2025-07-28, and Gemini-2.5-flash-lite-001) using a zero-shot prompting strategy with a unified evaluation schema (Sánchez-Soriano et al., 2026). The models received only the question text, the student's response, and the instructor-defined ideal answer; they were blind to human grades and other students' responses at all times.

Each LLM produces a structured JSON output containing four evaluation dimensions: coverage_accuracy (degree to which the answer covers the key concepts), justification_reasoning (quality of reasoning and explanation), terminology_clarity (correctness and clarity of technical terminology), and learning_transfer (whether the student demonstrates conceptual transfer). The output also includes a computed total_score on a 0 to 10 scale and a textual comment providing qualitative feedback. The unified JSON structure across models facilitates systematic comparison. A comprehensive description of the JorGPT system architecture, the automated evaluation protocol, and the data validation procedures is provided in Cisneros-González et al. (2025) and Sánchez-Soriano et al. (2026).

| Statistic | Value |
|---|---|
| Total number of student answers | 3041 |
| Number of distinct questions | 50 |
| Number of anonymized students | 79 |
| Courses/academic years | 2025–2026 |
| Average answer length (characters) | 542.43 |
| Grading scale | 0–10 |
| Number of LLM evaluators | 3: DeepSeek (deepseek-chat-V3.2), Qwen (qwen-flash-2025-07-28), Gemini (gemini-2.5-flash-lite-001) |
| Number of Judges (meta evaluator) | 1: OpenAI (gpt-4o-mini-2024-07-18) |
| Languages | Spanish (original), English (translation) |

*Figure 1.* Summary statistics of the JorGPT dataset (reproduced from Sánchez-Soriano et al., 2026).

To enable topic-level analysis, the 50 questions were classified into six knowledge domains based on their content and the corresponding ideal answers: TDD & Testing (10 questions, n = 758 responses), Docker & DevOps (10, n = 686), Vaadin & Frontend (12, n = 807), Spring Boot & REST (5, n = 226), JPA & Database (6, n = 263), and Architecture & Integration (7, n = 301). Classification was performed by the authors through independent review of the question content and ideal answer for each item.

## *2.2 Analysis Approach*

Our analysis combines quantitative and qualitative methods, proceeding from overall grading alignment to the internal structure of scores and then to the content and tone of feedback.

We first assess the alignment between each LLM's total score and the human instructor grade using Pearson correlation coefficient (r), Mean Absolute Error (MAE), and mean signed bias. Pearson r measures the strength of the linear relationship between the two sets of scores, MAE quantifies the average magnitude of scoring discrepancies, and mean bias captures the direction of systematic over- or under-grading. These metrics characterize the degree of agreement between two sets of ratings rather than establishing a causal relationship. To examine whether grading alignment varies by content domain, we compute these metrics within each of the six knowledge domains and apply the Kruskal-Wallis test, followed by pairwise Mann-Whitney U tests with Bonferroni correction.

We then examine the internal structure of LLM sub-dimension scores. For each LLM, we fit an ordinary least squares (OLS) regression model with the three sub-dimension scores as independent variables and the human instructor grade as the dependent variable. The learning_transfer dimension was excluded from the regression analysis because it operates as a binary indicator rather than a continuous scale. This regression examines the predictive contribution of each dimension and reveals the degree of redundancy among them. We assess multicollinearity via Variance Inflation Factors (VIF) and additionally report bivariate Pearson correlations between each individual sub-dimension and the human grade, which are not affected by inter-predictor correlations.

Finally, we analyze the qualitative feedback through two complementary approaches. First, we classify each feedback comment according to the presence or absence of four semantic roles: praise (acknowledging correct aspects), missing content identification (pointing out omitted information), misconception detection (flagging errors in understanding), and improvement suggestion (offering guidance for revision). Classification is performed using rule-based pattern matching with regular expressions targeting characteristic linguistic markers of each role (e.g., "you cover" and "correctly" for praise; "lack" and "missing" for missing content; "confus" and "incorrect" for misconception; "improve" and "should" for suggestion). Each comment may be assigned multiple roles, and we validated the approach by manually inspecting a stratified sample. Second, we analyze feedback tone using VADER (Hutto & Gilbert, 2014), a lexicon-based sentiment analysis tool that produces a compound score ranging from −1 to +1. We compare sentiment distributions between human and LLM feedback, stratified by student grade level (0 to 2, 3 to 5, 6 to 7, 8 to 10), to assess whether tone is appropriately modulated in response to student performance. As a supplementary indicator, we compare feedback length (word count) across sources.

## 3. Results

### *3.1 Overall Grading Alignment and Domain Variation*

Table 1 presents the overall grading alignment of each LLM with human instructor grades. All three models achieve moderate-to-strong correlations (r = 0.785–0.815), consistent with the range reported in prior studies of LLM-based short-answer grading (Cisneros-González et al., 2025; Khalf et al., 2026). The three models differ substantially in systematic bias (Friedman $\chi^2$ = 3,890.1 on signed errors, $p < .001$). DeepSeek exhibits systematic under-grading (bias = –0.55), Qwen shows pronounced leniency (bias = +1.28), and Gemini achieves near-zero bias (+0.06) with the lowest MAE (1.22).

Table 1. *Overall grading alignment of three LLMs with human instructor grades (N = 3,041).*

| Model | Pearson r | MAE | Mean Bias |
|---|---|---|---|
| DeepSeek-V3.2 | 0.813 | 1.39 | –0.55 (strict) |
| Qwen-flash | 0.785 | 1.61 | +1.28 (lenient) |
| Gemini-2.5-flash-lite | 0.815 | 1.22 | +0.06 (neutral) |

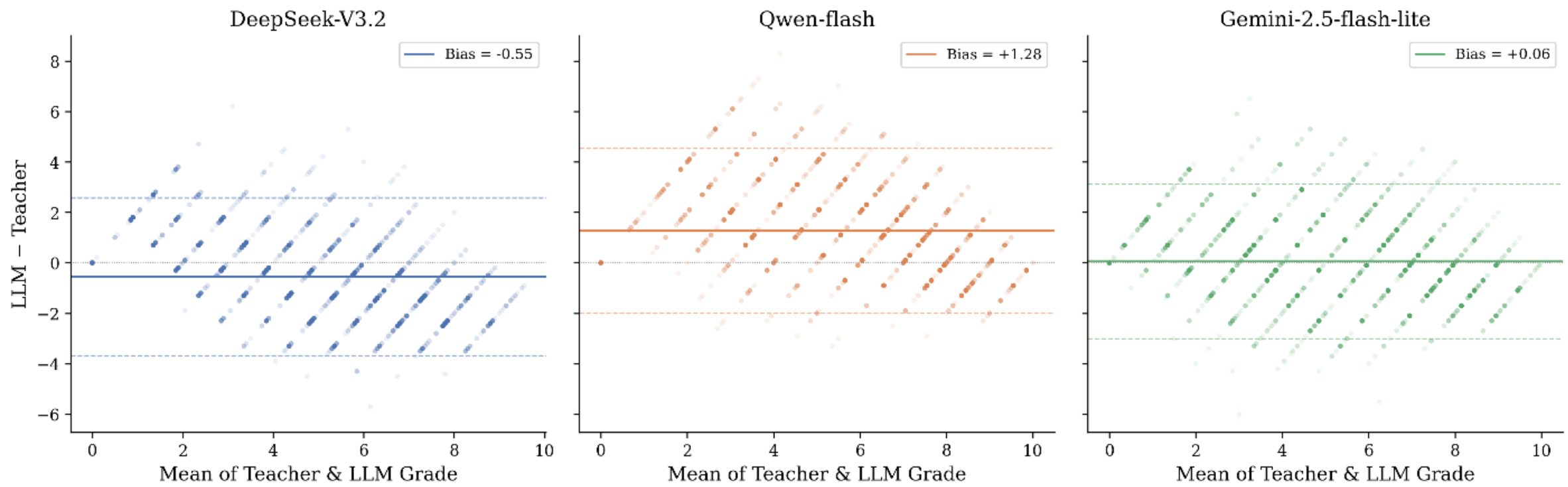

*Figure 2.* Bland-Altman plots showing systematic bias for each LLM.

Grading alignment varies significantly across knowledge domains (Kruskal-Wallis $p < 10^{-6}$ for all three LLMs). TDD & Testing is the most reliably graded topic across all models (MAE = 1.05–1.35, r = 0.84–0.87), significantly outperforming most other topics in post-hoc comparisons. Architecture & Integration and Spring Boot & REST yield the highest errors, with Qwen's Spring Boot performance particularly poor (MAE = 2.36, bias = +2.27). Bias direction also varies within models across topics: DeepSeek grades Architecture strictly (bias = –1.16) but Spring Boot leniently (+0.32). This pattern suggests that procedural and methodological knowledge, such as TDD testing cycles, aligns well with LLM evaluation capabilities, whereas topics requiring design reasoning are less reliably assessed.

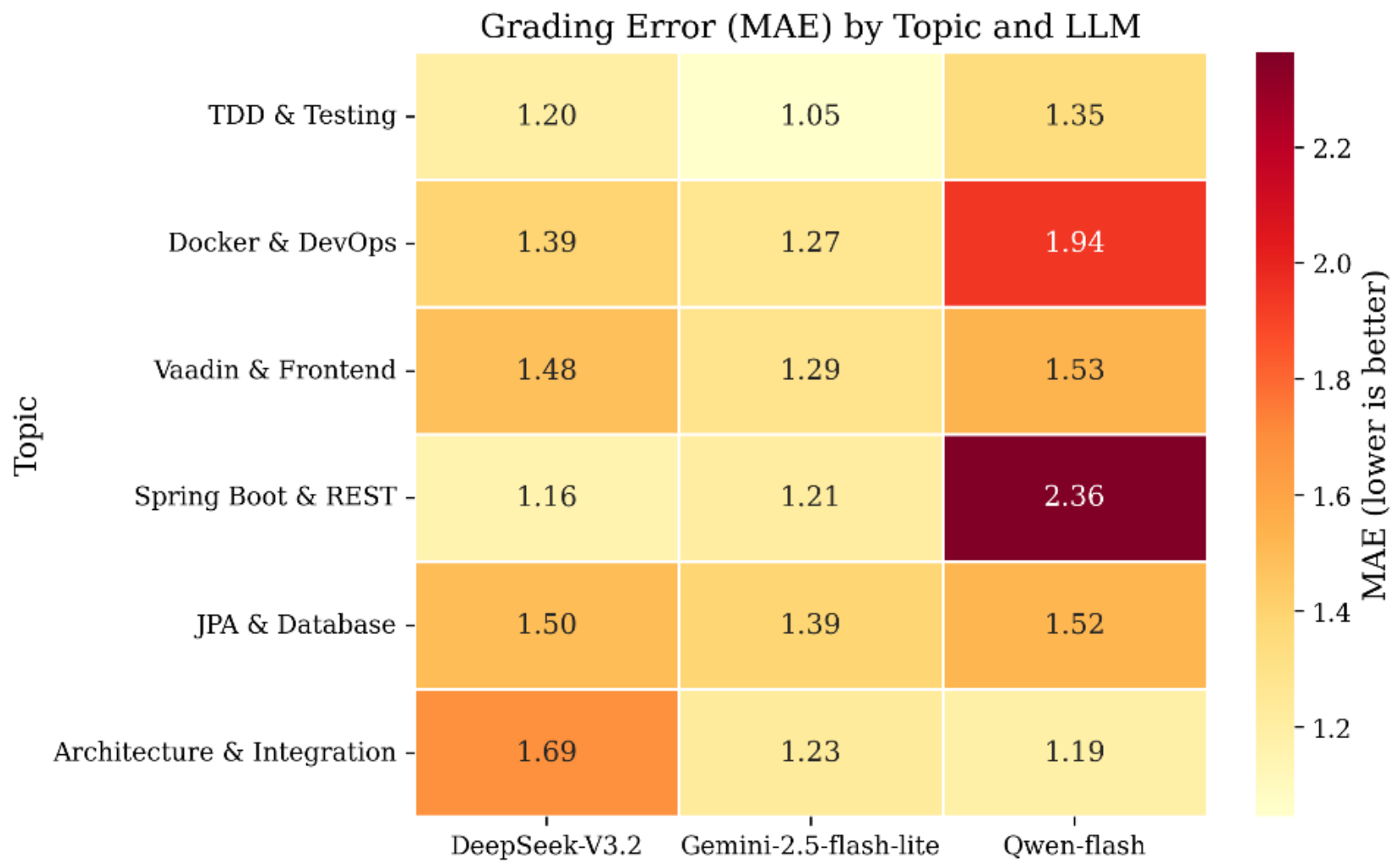

*Figure 3.* Heatmap of Mean Absolute Error by knowledge domain and LLM.

### *3.2 Independence of Sub-Dimension Scores*

Table 2 presents the OLS regression results for each LLM's sub-dimension scores predicting human instructor grades.

Table 2. *Regression coefficients (β), VIF, and bivariate correlations for sub-dimensions predicting human grades.*

| Dimension | DS β | VIF | Qw β | VIF | Ge β | VIF | Bivariate r |
|---|---|---|---|---|---|---|---|
| Coverage & Accuracy | 1.08*** | 15.1 | 0.75*** | 44.6 | 0.61*** | 17.3 | 0.78–0.80 |
| Justif. & Reasoning | 0.18** | 14.1 | 0.38*** | 41.1 | 0.31*** | 14.6 | 0.78–0.79 |
| Terminology & Clarity | –0.12** | 3.4 | 0.02 n.s. | 4.4 | 0.16*** | 7.1 | 0.65–0.76 |

*Note: $R^2$ = 0.64 (DeepSeek), 0.61 (Qwen), 0.64 (Gemini). *** $p < 0.001$, ** $p < 0.01$.*

Coverage & Accuracy is the dominant predictor across all models (β = 0.61–1.08, all $p < 0.001$). However, the VIF values reveal severe multicollinearity among the dimensions. Qwen's Coverage and Justification VIFs reach 44.6 and 41.1 respectively, far exceeding the conventional threshold of 5. The pairwise correlations between dimensions range from 0.82 to 0.99.

The consequences of this collinearity are evident in the regression coefficients. DeepSeek's Terminology coefficient is significantly negative (β = –0.12, p = 0.001), despite Terminology having a positive bivariate correlation with human grades (r = 0.65). This pattern is consistent with a suppression effect (MacKinnon et al., 2000): once Coverage and Justification absorb most of the shared variance, the residual variation in Terminology contributes negatively to the model. Qwen's Terminology coefficient is non-significant (p = 0.67), reflecting the same redundancy. These results indicate that the three sub-dimension scores, despite being presented as independent analytic assessments in a structured JSON format, largely capture a single underlying construct. The structured format may create unwarranted confidence in the diagnostic specificity of the grading.

### *3.3 Feedback Content and Tone*

*Semantic role analysis.* Table 3 presents the distribution of semantic roles identified in human instructor and LLM feedback.

Table 3. *Percentage of feedback comments containing each semantic role.*

| Semantic role | Teacher | DeepSeek | Qwen | Gemini |
|---|---|---|---|---|
| Praise | 63.7% | 83.5% | 87.1% | 78.1% |
| Missing content | 77.7% | 88.5% | 60.3% | 63.1% |
| Misconception | 15.0% | 7.4% | 4.8% | 6.0% |
| Suggestion | 15.8% | 92.0% | 89.7% | 74.6% |

The most notable difference appears in misconception detection. Human instructors flag conceptual errors in 15.0% of their feedback, compared with 4.8 to 7.4% for LLMs. This gap widens for low-performing students: among responses receiving human grades of 0 to 2, instructors identify misconceptions in 31.5% of their feedback, while LLMs do so in only 5 to 7%. Rather than diagnosing errors in understanding, LLMs default to identifying missing content and offering generic improvement suggestions.

The praise distribution further illustrates this pattern. For students scoring 0 to 2, LLMs include praise in 47 to 57% of their feedback, compared with 22.6% for human instructors. Similarly, LLMs offer improvement suggestions in 60 to 77% of low-scoring feedback, whereas teachers do so in only 3.3%. These patterns suggest that LLM feedback operates primarily as a coverage checklist, identifying what information was or was not mentioned, rather than as a diagnostic tool that assesses the nature and source of student errors.

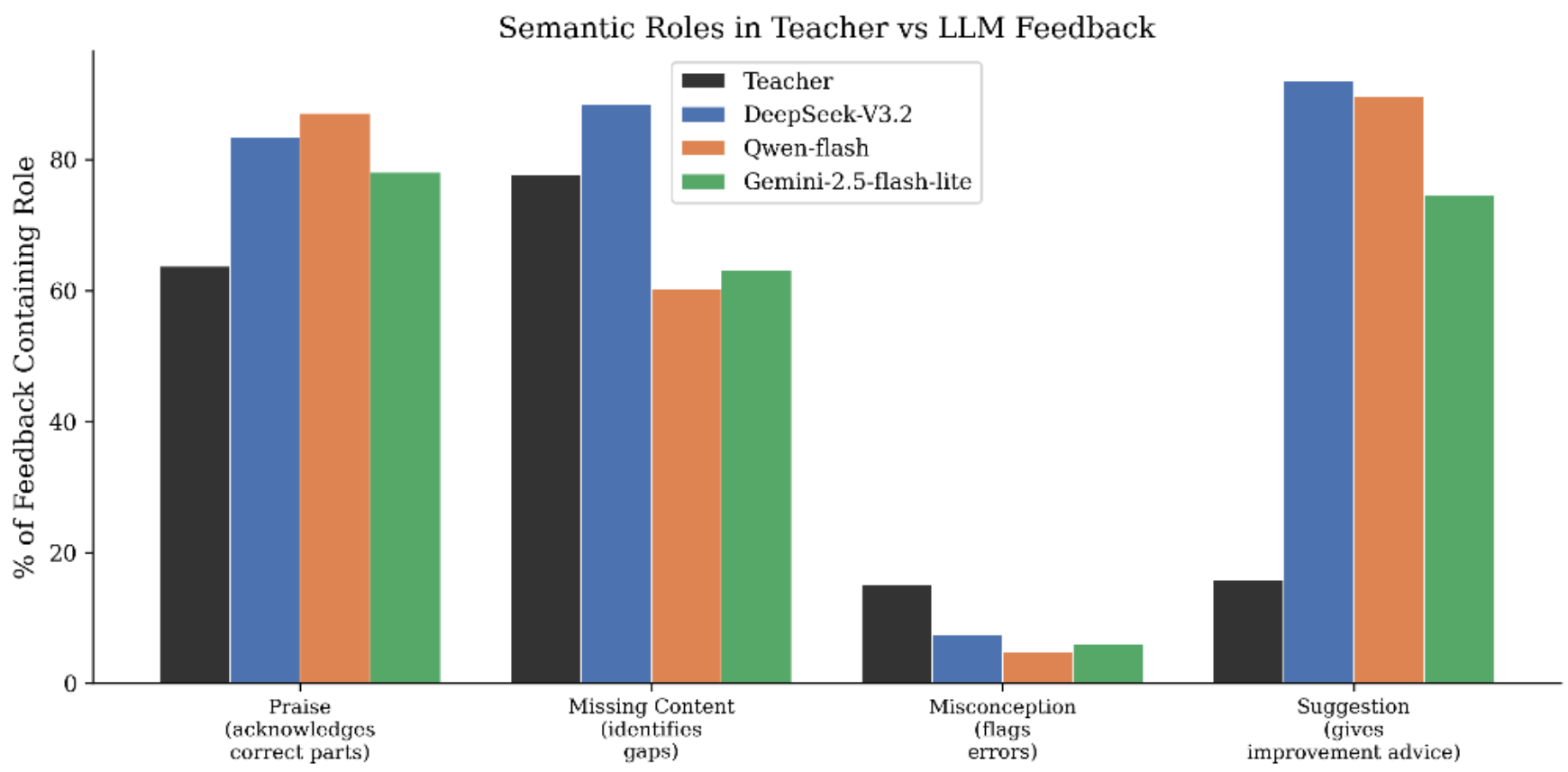

*Figure 4.* Semantic role distribution in teacher versus LLM feedback.

*Sentiment analysis.* VADER sentiment analysis reveals a systematic difference in how human instructors and LLMs modulate feedback tone. Human instructor feedback has a mean sentiment of 0.14, with 32.9% of comments classified as negative, 10.7% as neutral, and 56.3% as positive. LLM feedback averages 0.37 to 0.40, with over 80% classified as positive.

The grade-level breakdown (Figure 5) makes this divergence concrete. For students scoring 0 to 2, human instructor sentiment averages –0.07, providing a tonal signal that the response is seriously inadequate. For the same students, LLM sentiment ranges from +0.17 to +0.26, remaining in positive territory. Human instructor sentiment increases monotonically with student performance, reaching +0.45 for grades 8 to 10, creating a tonal gradient that may help students calibrate their self-assessment. LLM sentiment shows minimal variation across grade levels.

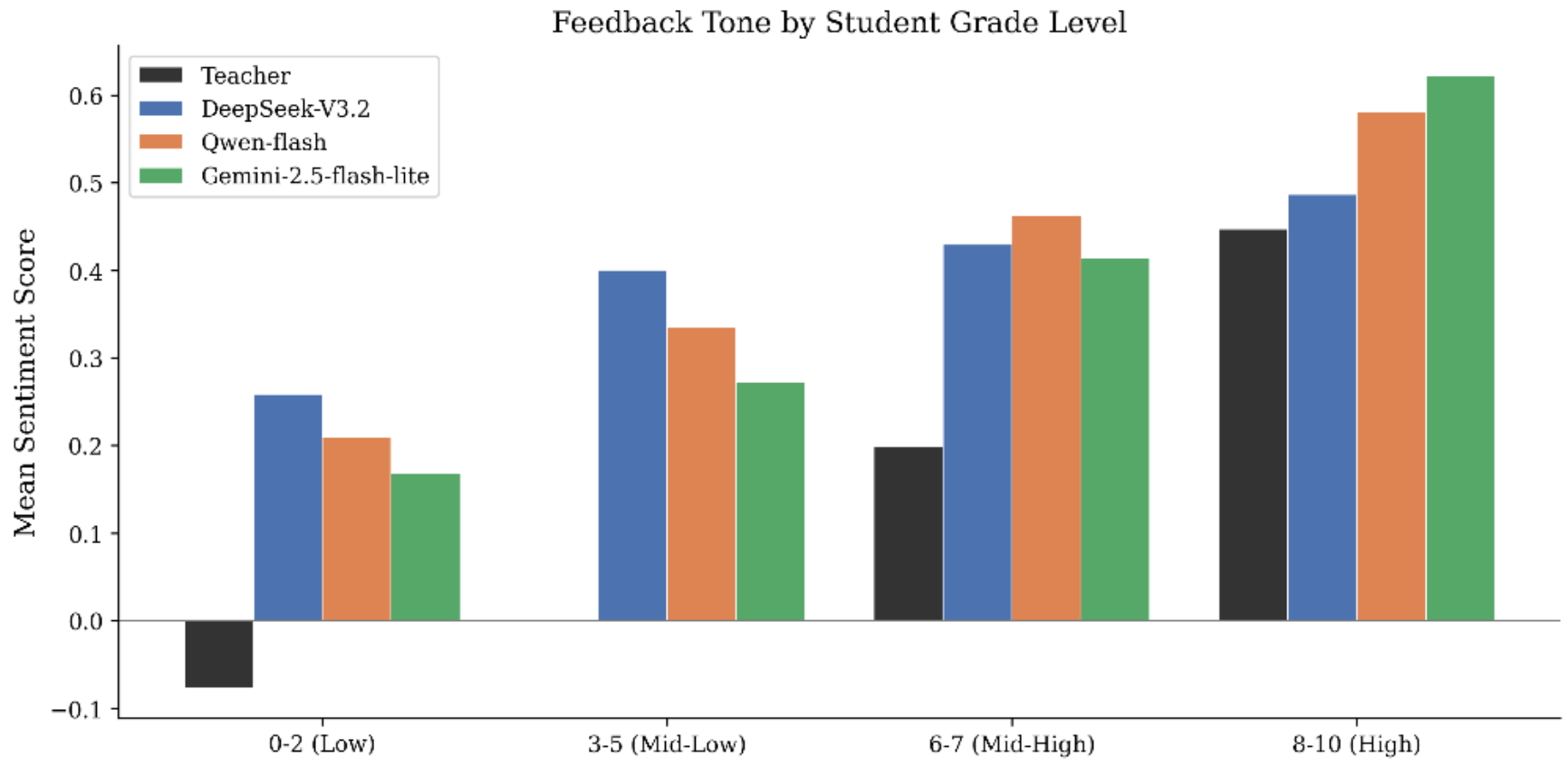

*Figure 5.* Feedback tone by student grade level.

Feedback length exhibits a complementary pattern. Human instructor feedback averages 59 words (SD = 22.8, range 2 to 314), while LLM feedback averages 23 to 28 words (SD = 6 to 12). Teacher feedback length correlates positively with student grade ($r = 0.17$, $p < 10^{-20}$), suggesting that instructors invest more detailed commentary in higher-quality responses. LLM feedback length shows no meaningful relationship with student performance.

## 4. Discussion

This study examined the quality of LLM-generated grading and feedback in an undergraduate software engineering course, moving beyond aggregate score accuracy to investigate whether the diagnostic and formative information embedded in structured outputs matches their apparent sophistication. Our analyses reveal three systematic discrepancies that carry practical implications for how educators interpret and use LLM-based grading and feedback tools.

At the score level, all three LLMs achieve moderate-to-strong correlations with human grades ($r = 0.785$ to $0.815$), consistent with the range reported in prior work (Khalf et al., 2026). However, the substantial variation in systematic bias across models, combined with the significant domain-level differences in grading accuracy, suggests that model selection and content domain are critical considerations for educators adopting LLM-based assessment. The finding that procedural topics such as TDD are more reliably assessed than design-thinking topics such as architecture is consistent with the broader observation that LLMs perform better on tasks with well-defined correctness criteria (Gao et al., 2024). For educators, this implies that LLM-based assessment may be most appropriate as a screening tool in procedural knowledge domains, while topics requiring evaluative judgment should continue to receive human assessment.

Looking beyond aggregate scores, the high multicollinearity among sub-dimension scores (VIF up to 45, inter-dimension $r$ up to 0.99) indicates that the structured format of LLM outputs can create unwarranted confidence in diagnostic specificity. Multi-dimensional rubric scores are traditionally intended to provide independent feedback on distinct aspects of student performance (Jonsson & Svingby, 2007), but the LLM-generated dimensions largely capture a single underlying construct. Recent work on LLM-based evaluation suggests that this may reflect a broader pattern. W. Li et al. (2025) identified a "criteria entanglement bias" in which LLM judges presented with multiple evaluation criteria tend to assign similar scores across dimensions, producing a halo effect that undermines the intended granularity of analytic rubrics. Kucia et al. (2026) reported a complementary finding: LLMs achieve higher agreement with human raters under holistic scoring rubrics than analytic ones, suggesting that these models process responses through shared representations rather than evaluating each dimension independently. These findings suggest that educators should not rely on individual sub-dimension scores as indicators of specific strengths or weaknesses in student understanding. The practical implication is that the structured JSON format, while visually resembling an analytic rubric, functions more like a holistic assessment presented in analytic clothing.

The qualitative analysis of feedback content reveals a more fundamental limitation. LLMs are substantially less likely than human instructors to detect student misconceptions (5 to 7% vs. 15 to 31% in low-scoring responses), instead defaulting to coverage-based evaluation. This pattern is consistent with the finding that chatbot-generated feedback can include irrelevant or superficial suggestions (Usher, 2025). For formative assessment, where identifying the nature of student errors is considered more valuable than simply indicating their presence (Hattie & Timperley, 2007), this represents a significant gap. Educators relying on LLM feedback may receive accurate information about what students omitted but not about what students misunderstood. The uniformly positive tone of LLM feedback, regardless of response quality, compounds this concern. In educational contexts, tone serves as an implicit indicator of severity that helps students calibrate their self-assessment. The absence of tonal modulation in LLM feedback is consistent with research showing that LLM-generated responses tend toward emotional uniformity in educational settings (Liu et al., 2025), and it may reduce the communicative effectiveness of the feedback for students who most need to recognize the inadequacy of their responses.

Several limitations of this study should be acknowledged. The data come from a single institution and course in software engineering, which constrains the generalizability of the findings to other disciplines and educational contexts. The rule-based semantic role classification captures prominent linguistic patterns but may miss subtle or atypical expressions, and VADER sentiment analysis, while established for short texts, may not fully capture the nuanced tone of educational feedback. The three LLMs studied represent a specific generation of models, and newer models may exhibit different behavioral patterns. Future research should validate these findings across disciplines, languages, and model versions. It would also be valuable to investigate whether prompt engineering or fine-tuning strategies can improve the diagnostic specificity of LLM feedback, and whether explicitly instructing models to detect misconceptions or modulate tone yields measurable improvements in feedback quality.